# Symbiotic Stars as Laboratories for the Study of Accretion and Jets: A Call for Optical Monitoring

## J. L. Sokoloski

*Harvard-Smithsonian Center for Astrophysics, 60 Garden Street, Cambridge, MA 02138*

**Abstract**   Symbiotic binary stars typically consist of a white dwarf (WD) that accretes material from the wind of a companion red giant. Orbital periods for these binaries are on the order of years, and their relatively small optical outbursts tend to occur every few years to decades. In some symbiotics, material that is transferred from the red giant to the WD forms a disk around the WD. Thus, symbiotic stars are a bit like overgrown cataclysmic variables (CVs), but with less violent eruptions. Symbiotic stars are not as well understood as CVs, in part because their longer variability time scales mean that observations over many years are required to cover different outburst states and orbital phases. The recent discovery of collimated outflows ("jets") from a number of symbiotics provides a new motivation for such long-term study of these objects. Astrophysical jets are observed in almost every type of accretion-powered system, and symbiotic stars may help us understand these structures. In symbiotics, most jets appear to be associated with optical eruptions. Optical monitoring by amateurs can identify systems in outburst, and also help to build a comprehensive database of outburst and quiescent symbiotic light curves. Together with radio through X-ray observations that will be performed when new outbursts are found, long-term optical light curves will improve understanding of symbiotic outbursts, jet production, and the connection between outbursts, jets, and accretion disks in symbiotic stars.

### 1. Introduction: What are symbiotic stars?

Symbiotic stars are interacting binary stars in which a hot white dwarf (WD; or in a few cases a main-sequence star) orbits a red-giant star and captures material from the wind of the red giant (RG). A schematic diagram of the main components of a symbiotic binary is shown in Figure 1. Because a WD is compact (roughly the size of the earth, but two hundred thousand times more massive), it has a strong gravitational field and can capture more of the wind than would otherwise hit the WD directly. As the accreted wind material falls toward the WD, it is accelerated and heated, and energy is released. In some symbiotic binaries, the accreted material may form a disk around the WD. Figure 2 is a computer-generated picture (not to scale) of the two stars, including an accretion disk and a bi-polar (*i.e.*, two-sided) jet (discussed in Section 2). Symbiotic stars are thus similar to cataclysmic variable stars (CVs). However, whereas a typical CV binary is not much larger than our sun, the distance between the stellar components in a standard symbiotic binary is comparable to the distance between the earth and the sun.

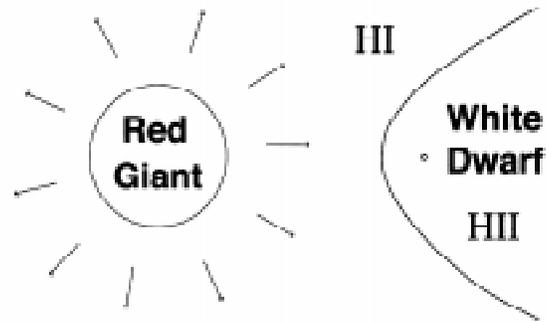

Figure 1. Schematic drawing of the main components of a symbiotic binary: the hot white dwarf (small circle on the right of the diagram), the cool red giant losing material in a wind, and the partially ionized nebula. In this picture, the bow-shaped curve represents the boundary between the portion of the nebula that is ionized by radiation from the hot WD (ionized hydrogen is designated HII), and the region that is primarily neutral hydrogen (HI). In different systems, this boundary can have different shapes.

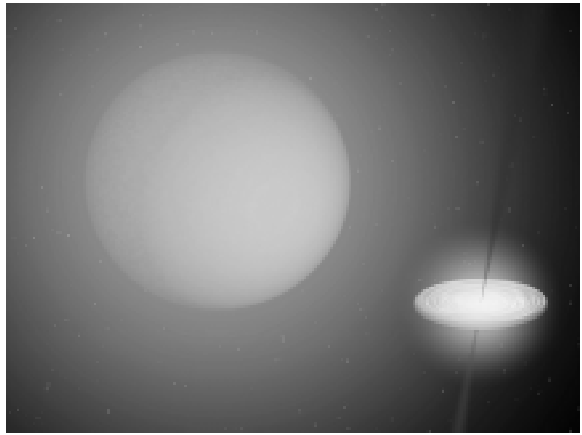

Figure 2. This computer-generated picture shows the components believed to comprise a jet-producing symbiotic binary (except the extended nebula): a large red-giant star losing material in a wind, and a compact white dwarf surrounded by an accretion disk. The accreting WD occasionally produces a collimated jet. *Image courtesy R. Hynes.*

Symbiotic stars are also closely related to the other main type of accreting-white-dwarf binary: supersoft X-ray sources. If the accretion rate onto the white dwarf in a symbiotic binary is within a certain range, the temperature and pressure at the base of the layer of accreted hydrogen-rich material will be very high. The hydrogen may then undergo quasi-steady nuclear fusion burning (Paczyński and Żytków 1978; Fujimoto 1982), causing additional energy to be released. Optical, UV, and X-ray spectra, as well as the reduced amplitude of rapid optical flickering from an accretion disk (usually so low it is undetectable), compared to that in CVs, support the idea that nuclear burning on the surface of the white dwarf is the dominant source of power in most symbiotics (*e.g.*, Mürset *et al.* 1991; Sokoloski, Bildsten, and Ho 2001). Such quasi-steady nuclear burning on the surface of a white dwarf is also thought to power supersoft X-ray sources (van den Heuvel *et al.* 1992).

Most of the wind from the red giant in a symbiotic binary is not accreted onto the WD. Instead, it forms a large nebula around the binary that is partially ionized by radiation from the accreting WD and the nuclear fusion on its surface. The ionized nebula radiates in the optical, and thereby conveys information about the X-ray- and far-ultraviolet-emitting WD into the optical regime. Symbiotic optical spectra thus show both absorption lines from the surface of the cool red giant, and emission lines from the nebula that require the presence of a hot source of ionizing radiation. These combination spectra are the reason symbiotic stars were originally called "symbiotic". Spectral features indicating the presence of two very different types of environments in the same object made these objects seem a symbiosis of hot and cold that we now know comes from two separate stars in a binary.

**2. Jets**

Collimated outflows of material ("jets") have been observed from at least 10 of roughly 200 known symbiotic stars (see Brocksopp *et al.* 2003 for a list and references). Most of these jets are transient, and seem to appear during or after an optical outburst, or temporary brightening. They then tend to fade on time scales of months to years. Therefore, more jets are likely to be found when other symbiotic stars go into outburst. Symbiotic jets are small in the angle they subtend on the sky (typically a few arcsec or less; see Figure 3), so they are a challenge to observe. They can either be detected directly with high-angular-resolution radio maps, or their presence inferred from Doppler-shifted emission lines in optical spectra. Moreover, the radio observations or optical spectroscopic observations must be performed at the correct time, after material has been ejected in the form of a jet, but before emission from the jet fades.

Astrophysical jets are seen in systems ranging from black holes in X-ray binaries, to pre-main-sequence stars, to active galactic nuclei. How these jets are produced is an important unsolved problem. As in the other astrophysical jet sources, it is not fully understood how symbiotic-star jets are collimated and accelerated. The outflows are somehow linked to the process of accretion. For example, an accretion disk threaded by a magnetic field, plus some additional source of energy or wind close to the central object, may be required for jets to form (Livio 1997). If the basic ingredients for jet collimation and acceleration are the same in all cases, then the detailed investigation of one class of jet source promises to improve our fundamental understanding of the production of astrophysical jets. Symbiotic stars are likely to have accretion disks that are similar to the disks in CVs, which are among the best understood disks. In addition, symbiotics are numerous, and they are close enough for their jets to be imaged (see Figure 3). They are therefore good objects to study. Furthermore, some of the main elements apparently necessary for the generation of jets (*e.g.*, an accretion disk, a strong white-dwarf magnetic field, and nuclear burning on the WD surface) exist in some symbiotic stars but not in others. *[Author's note: Mikołajewska (2003) has suggested that the symbiotics with accretion disks are the ones that have outbursts, and Livio (1988) described requirements for disk formation that can be used to determine which systems would be expected to contain accretion disks on theoretical grounds. For a variety of perspectives on magnetic fields in symbiotic stars and a report of the first solid detection of a strong magnetic field in a symbiotic, see Sokoloski (2003); Sokoloski, Bildsten, and Ho (2001); Sokoloski and Bildsten (1999); Mikołajewski, Mikołajewska, and Khudyakova (1990); and Tomov (2003). Finally, the luminosity of the white dwarf in a symbiotic indicates whether or not extra energy is being released by quasi-steady nuclear burning on the WD surface; estimates of these luminosities are reported by Mürset et al. (1991).]* Thus, the necessity of these elements for jet production can in principle be tested.

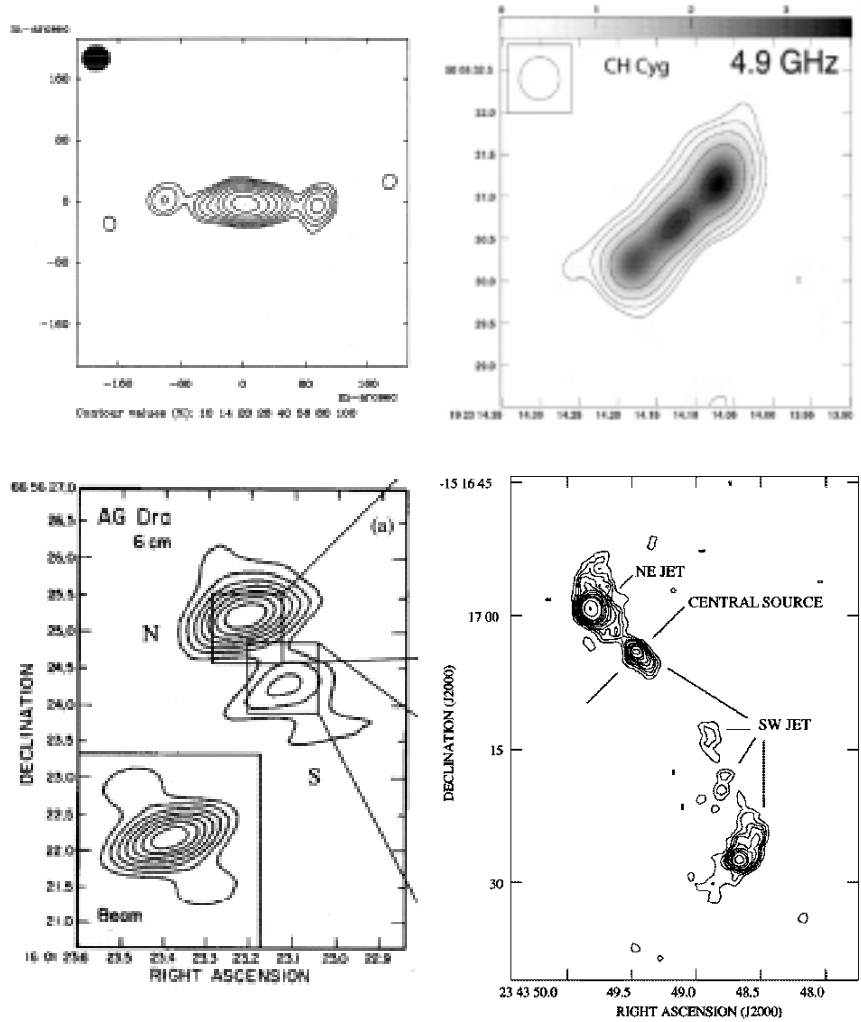

Figure 3: Four examples of symbiotic-star radio jets. Upper left: RS Oph (Taylor *et al.* 1989). Upper right: CH Cyg (Crocker *et al.* 2001). Lower left: AG Dra (Ogley *et al.* 2002). Lower right: R Aqr—the X-ray contour map is shown for this system (Kellogg *et al.* 2001). All graphs used by permission of the authors.

### 3. Outbursts

Most symbiotic jets occur together with an optical outburst or change of brightness state. Therefore, investigations of jets rely on awareness of symbiotic outbursts. Moreover, understanding jet production requires knowledge of the cause of the outbursts. There are at least three types of outbursts in symbiotic stars: recurrent novae, symbiotic (or "slow") novae, and "classical symbiotic outbursts." The first two types of outbursts are believed to be due to a thermonuclear runaway on the surface of the white dwarf, as in classical novae. However, the cause of classical symbiotic outbursts, which are the most common, is not known. During a classical symbiotic outburst, the optical brightness typically increases by several magnitudes, but in some cases can increase by as little as one magnitude. The system may take weeks or months to reach maximum brightness, and then months or years to fade. These events could be due to: (a) gravitational energy release from a sudden influx of matter onto the WD due to an accretion disk instability (as in the dwarf novae CVs; see Warner 1995); (b) an expansion of the WD photosphere, shifting the high luminosity of a steady-burning WD from the UV into the visible part of the spectrum; (c) an increase in the rate of nuclear burning on the WD surface at the base of its accreted surface layers; or some combination of all of these phenomena.

The nature of symbiotic-star outbursts may also have bearing on cosmological studies of the expansion of the universe. The amount of inflow to and outflow from a WD in a symbiotic binary determines whether it can gain enough material to approach the maximum allowable mass for a WD (the Chandrasekhar limit) and then explode as a Type Ia supernova (SN Ia). SN Ia are used as "standard candles" to determine distances in cosmological studies. It is therefore important to identify their progenitor objects so that any intrinsic differences between SN Ia at different cosmic epochs can be fully taken into account. To explode as a SN Ia, the mass of an accreting WD must eventually approach the Chandrasekhar mass limit of 1.4 solar masses. But the amount of material ejected from a symbiotic WD during outburst (in the form of a jet or spherical shell) limits the degree to which the WD mass can increase during the period of active mass transfer from the red giant. Therefore, understanding the outburst mechanism(s) will help reveal the fate of symbiotic WDs, and their relation to SN Ia.

**4. Observing strategy**

Multi-wavelength and "Target-of-Opportunity" (TOO) observing strategies are necessary to investigate symbiotic-star outbursts and jets. The bulk of the energy produced by the hot WD in symbiotic stars is radiated at short wavelengths (primarily far ultraviolet and soft X-rays), whereas any disk emission is likely to be optical, and much of the emission from the nebula is in the optical and the radio. Furthermore, collimated jets can often only be directly detected at radio wavelengths. Therefore, observations over a broad range of wavebands are required for imaging and spectral modeling.

Target-of-Opportunity observations are planned ahead of time, and performed when a certain trigger event, such as an outburst, occurs. TOO observations are needed to study symbiotic stars because their outbursts are not predictable. TOO programs are in place for observations with the Very Large Array (VLA; radio), the Hubble Space Telescope (HST), the Far Ultraviolet Spectroscopic Explorer (FUSE) satellite, the X-ray Multi-Mirror (XMM) satellite, and the Chandra X-ray Observatory, in addition to optical spectroscopy. However, all of these potentially important observations rely on knowing when a symbiotic star is in outburst.

**5. Role of optical monitoring**

Optical monitoring is important for many aspects of symbiotic star research. Long-term light curves will: 1) reveal systems in outburst; 2) enable TOO observations with the VLA, as well as HST, FUSE, and in some cases XMM and Chandra satellites; 3) determine the fraction of symbiotic stars that have outbursts; 4) determine other basic outburst statistics, such as average outburst frequencies and durations; and

5) generally characterize symbiotic-star long-term optical variability (classical symbiotic outbursts may actually encompass several different types of outbursts). Table 1 lists the symbiotic stars that are already included in the AAVSO International Database. Observations of objects from this list are welcomed and encouraged. We are working to expand the set of AAVSO symbiotics to include all symbiotic stars listed in the catalog of Belczyński *et al.* (2000), hopefully within the next year or two. Table 2 lists objects that have not yet been added to the AAVSO database, but for which observations are needed for an X-ray program. Observers interested in these objects should contact the author directly for charts (jsokolos@cfa.harvard.edu).

## 6. Summary

Symbiotic stars are wide binary stars in which material is transferred to a compact white dwarf from an evolved red giant. Because of the longer variability time scale compared to CVs and the presence of a luminous nebula around the two component stars, the details of the accretion process in symbiotic stars have been more difficult to ascertain than in CVs. Thus, many interesting and fundamental open questions remain. For example, symbiotic stars appear to be a new class of jet-producing astronomical objects, but how are these jets produced? In addition, the cause of the most common type of symbiotic-star outburst is still not understood, and symbiotics may be the progenitors of the cosmologically important Type Ia supernovae. The potential value of comprehensive long-term optical monitoring of symbiotic stars provides an exciting opportunity for groups of amateurs such as the AAVSO and their professional collaborators. With enough amateur observations: 1) all outbursts can be discovered and studied, 2) outburst statistics for the whole class can be obtained and outburst classification refined, and finally, 3) the relationship between outbursts and jets can be revealed. A new program is being initiated to increase the number of symbiotic stars in the AAVSO database from roughly 70 to the full set of approximately 200. A workshop will be held at a future AAVSO meeting for those interested in learning more about symbiotic stars and how to observe them. Since long-term optical monitoring is both valuable in its own right and can lay the groundwork for additional observations, collaboration between amateurs and professionals is an excellent way to make progress in this field of study.

## 7. Acknowledgements

Thanks to P. Maragakis and J. Aufdenberg for assistance with Figure 1, and S. Kenyon, R. Webbink, and J. McDowell for helpful comments. I am also grateful to M. Saladyga for valuable assistance with the tables and Figure 3.

**8. Further reading**

Corradi, R. L. M., Miko ajewska, J., and Mahoney, T. J., eds. 2003, *Symbiotic Stars Probing Stellar Evolution*, ASP Conference Series, Vol. 303, San Francisco, Astronomical Society of the Pacific.

Kenyon, S. J. 1986, *The Symbiotic Stars*, Cambridge U. Press, Cambridge and New York.

Livio, M. 1997, in *Accretion Phenomena and Related Outflows*, IAU Colloquium 163, ASP Conference Series Vol. 121, D. T. Wickramasinghe, L. Ferrario, and G. V. Bicknell, eds., p. 845.

Table 1. Symbiotic stars currently in the AAVSO International Database.

| Desig. | Name | R. A. (2000) h m s | Dec. (2000) ° ′ ″ | V | Coordinates Source / Comments |
|---|---|---|---|---|---|
| 0039+40 | EG And | 00 44 37.1799 | +40 40 45.838 | 7.1 | HIP 3494 |
| 0130+53 | AX Per | 01 36 22.6957 | +54 15 02.456 | 10.9 | TYC 3671 247 1 |
| 0152+52 | V471 Per | 01 58 49.6751 | +52 53 48.473 | 13.0 | CMC 901416 few data points |
| 0214–03 | omi Cet | 02 19 20.7866 | –02 48 37.418 | 6.0 | HIP 10826 Mira prototype |
| 0515+32 | UV Aur | 05 21 48.9243 | +32 30 40.105 | 8.5 | TYC 2394 373 1 |
| 0517–08 | V1261 Ori | 05 22 18.6169 | –08 39 57.986 | 6.8 | HIP 25092 few data points |
| 0720–03 | BX Mon | 07 25 22.7644 | –03 35 50.424 | 11.7 | CMC 905899 |
| 0721–07 | V694 Mon | 07 25 51.2856 | –07 44 08.111 | 9.5 | TYC 5396 1135 1 known jet source |
| 0810–41 | RX Pup | 08 14 12.3203 | –41 42 28.956 | 11.5 | TYC 7668 2915 1 |
| 0827–27 | AS 201 | 08 31 42.890 | –27 45 31.60 | 11.8 | USNO-A2.0 0600 -10606304 few data points |
| 0937–48 | KM Vel | 09 41 13.6529 | –49 23 27.798 | 15.0 | UCAC1 11036888 few data points |
| 1127–64A | SY Mus | 11 32 09.9890 | –65 25 11.582 | 10.9 | TYC 8980 154 1 |
| 1217–62 | BI Cru | 12 23 25.9959 | –62 38 16.012 | 11.1 | UCAC1 04287027 check out this light curve!! |
| 1229–64 | RT Cru | 12 34 53.7354 | –64 33 56.091 | 12.6 | UCAC1 03422435 few data points |
| 1239+37 | TX CVn | 12 44 42.0752 | +36 45 50.648 | 9.5 | TYC 2533 1168 1 |
| 1310–36 | V1044 Cen | 13 16 01.6 | –37 00 11.9 | 11.2 | A&AS 146, 407 CD–368436 =NSV 6160 |
| 1314–55 | V840 Cen | 13 20 49.401: | –55 50 14.50: | 14.1 | GSS [GSC 8666 -01230] |
| 1328–24 | RW Hya | 13 34 18.1438 | –25 22 48.987 | 8.9 | TYC 6718 1146 1 |
| 1328–64 | NSV 19878 | 13 35 27.5605 | –64 45 44.995 | 12.9 | UCAC1 03446120 = Hen 3-916; few data points |
| 1408–61 | V417 Cen | 14 15 59.6796 | –61 53 50.232 | 12.2 | UCAC1 04818754 |
| 1411–21 | IV Vir | 14 16 34.2881 | –21 45 49.920 | 10.7 | TYC 6151 1012 1 =BD-21 3873 |
| 1537–66 | Hen3-1092 | 15 47 10.6 | –66 29 16.0 | 13.5 | A&AS 146, 407 few data points |

Table 1. Symbiotic stars currently in the AAVSO International Database, continued.

| Desig. | Name | R. A. (2000) h m s | Dec. (2000) ° ' " | V | Coordinates Source / Comments |
|---|---|---|---|---|---|
| 1544–48 | NSV 20412 | 15 51 15.9327 | –48 44 58.529 | 11.0 | HIP 77662 = HD 330036; few data points |
| 1555+26 | T CrB | 15 59 30.1650 | +25 55 12.507 | 10.1 | HIP 78322 recurrent nova |
| 1601+67 | AG Dra | 16 01 41.0226 | +66 48 10.187 | 9.1 | HIP 78512 radio jet detected |
| 1635–62 | KX TrA | 16 44 39.7893 | –62 37 05.498 | 12.4 | UCAC1 04373496 |
| 1645–25 | NSV 20790 | 16 51 20.4061 | –26 00 26.782 | 12.2 | UCAC1 22568929 = AS 210; few data points |
| 1648–30C | HK Sco | 16 54 41.0412 | –30 23 06.848 | 13.5 | UCAC1 20442067 |
| 1648–30A | CL Sco | 16 54 51.9704 | –30 37 18.232 | 13.3 | UCAC1 20164154 |
| 1700–33 | V455 Sco | 17 07 21.7369 | –34 05 14.458 | 13.7 | UCAC1 18355984 few data points |
| 1702–17 | V2523 Oph | 17 08 36.6 | –17 26 30.0 | 12.5 | A&AS 146, 407 = Hen 3-1341 |
| 1734–11 | RT Ser | 17 39 51.9905 | –11 56 38.598 | 15.0 | CMC 1014993 slow nova; few data points |
| 1733–47 | AE Ara | 17 41 04.9178 | –47 03 27.197 | 12.5 | UCAC1 12127941 |
| 1737–22 | V2110 Oph | 17 43 33.366 | –22 45 35.91 | 19 | USNO-A2.0 0600 -28373042 slow nova |
| 1744–06 | RS Oph | 17 50 13.1626 | –06 42 28.512 | 11.5 | CMC 1110739 recurrent nova; radio jet |
| 1745–17 | ALS 2 | 17 50 51.1135 | –17 47 57.220 | 14.2 | UCAC1 26364569 few data points |
| 1745–22 | AS 245 | 17 51 00.860 | –22 19 35.45 | 11.0 | USNO-A2.0 0675 -22897072 few data points |
| 1748–33B | V745 Sco | 17 55 22.235 | –33 14 58.57 | 17 | GSS recurrent nova |
| 1759–20 | AS 270 | 18 05 33.7281 | –20 20 38.086 | 13.1 | UCAC1 25437415 outburst mid-2001? |
| 1800–36 | V615 Sgr | 18 07 39.922 | –36 06 22.13 | 13.3 | GSS outburst late 1997? |

Table 1. Symbiotic stars currently in the AAVSO International Database, continued.

| Desig. | Name | R. A. (2000)<br>h  m  s | Dec. (2000)<br>°  '  " | V | Coordinates Source /<br>Comments |
|---|---|---|---|---|---|
| 1804–28 | V2506 Sgr | 18 11 01.679 | –28 32 39.03 | 12.0 | GSS<br>few data points |
| 1806–11 | V343 Ser | 18 12 22.153 | –11 40 07.17 | 12.1 | USNO-A2.00750<br>-12799421<br>= AS 289; few data points |
| 1807–42 | Y CrA | 18 14 22.9523 | –42 50 32.296 | 14.4 | UCAC1 14276652 |
| 1810+20 | YY Her | 18 14 34.1747 | +20 59 21.084 | 12.8 | CMC 412559<br>finished outburst in June 2003 |
| 1808–29B | V2756 Sgr | 18 14 35.175 | –29 49 20.00 | 11.5 | GSS |
| 1809–00 | FG Ser | 18 15 07.0959 | –00 18 52.107 | 11.0 | TYC 5098 731 1 |
| 1809–30 | V4074 Sgr | 18 16 05.561: | –30 51 15.31: | 11.5 | GSS |
| 1811–28 | V2905 Sgr | 18 17 20.303 | –28 09 49.62 | 12.3 | GSC 6856-01061 |
| 1810–66 | AR Pav | 18 20 27.8823 | –66 04 42.878 | 10.0 | HIP 89886 |
| 1815–31 | V3804 Sgr | 18 21 28.7799 | –31 32 04.554 | 12.0 | UCAC1 19672940 |
| 1818+23 | V443 Her | 18 22 07.883 | +23 27 20.16 | 11.5 | A&AS 143, 343 |
| 1819–28 | V4018 Sgr | 18 25 27.0 | –28 35 57.5 | 11.4 | 2-mag drop started May 2000 |
| 1824–24 | V3890 Sgr | 18 30 43.2875 | –24 01 08.901 | ~14 | UCAC1 23612566<br>recurrent nova |
| 1842–20 | NSV 24592 | 18 47 55.8072 | –20 05 50.964 | 12.0 | UCAC1 25466410<br>= MWC 960;<br>few data points |
| 1847–24B | AS 327 | 18 53 16.6569 | –24 22 58.767 | 11.8 | UCAC1 23632440<br>few data points |
| 1848–19 | FN Sgr | 18 53 54.7850 | –18 59 40.668 | 12.7 | CMC 1111581<br>outburst ~1995–2001 |
| 1857–17 | V919 Sgr | 19 03 45.1332 | –16 59 55.291 | 12.2 | CMC 917107 |
| 1859+16 | V1413 Aql | 19 03 46.8385 | +16 26 17.110 | 14.0 | CMC 1017292<br>outburst ~1993–2001 |
| 1920+29 | BF Cyg | 19 23 53.5039 | +29 40 29.250 | 12.3 | TYC 2137 234 1 |
| 1921+50 | CH Cyg | 19 24 33.0742 | +50 14 29.301 | 7.1 | HIP 95413<br>known jet source |
| 1923–06 | StHA 164 | 19 28 41.653 | –06 03 51.54 | 13.6 | A&A 372, 145 |

Table 1. Symbiotic stars currently in the AAVSO International Database, continued.

| Desig. | Name | R. A. (2000) h m s | Dec. (2000) ° ' " | V | Coordinates Source / Comments |
|---|---|---|---|---|---|
| 1937+16 | HM Sge | 19 41 57.0688 | +16 44 39.710 | 17 | TYC 1602 1636 1; slow nova |
| 1932–68 | NSV 12264 | 19 42 25.3 | –68 07 35.3 | 10.4 | A&AS 146, 407 = Hen 3-1761; outburst 1999 |
| 1941+18 | QW Sge | 19 45 49.548 | +18 36 48.47 | 12.8 | A&AS 143, 343; few data points |
| 1946+35 | CI Cyg | 19 50 11.8360 | +35 41 03.028 | 11.0 | HIP 97594 |
| 1953+39 | V1016 Cyg | 19 57 05.0301 | +39 49 36.162 | 11.2 | TYC 3141 533 1; slow nova |
| 1956–56 | RR Tel | 20 04 18.5345 | –55 43 33.144 | 10.8 | TYC 8780 1277 1; slow nova |
| 2016+21 | PU Vul | 20 21 13.3167 | +21 34 18.133 | 11.6 | TYC 1643 1021 1; slow nova |
| 2031+19 | LT Del | 20 35 57.234 | +20 11 27.91 | 13.1 | A&AS 143, 343; few data points |
| 2037+08 | ER Del | 20 42 46.4962 | +08 41 13.438 | 10 | TYC 1089 194 1 |
| 2047+35 | V1329 Cyg | 20 51 01.279 | +35 34 54.59 | 13.3 | USNO-A2.0 1200 -1619252 1; slow nova; jet source |
| 2053–43 | CD–43 14304 | 21 00 06.3593 | –42 38 44.899 | 10.8 | UCAC1 14307392; observations greatly needed |
| 2058+45 | V407 Cyg | 21 02 09.831 | +45 46 32.85 | 14 | A&AS 143, 343; outburst ~1998–2003 |
| 2146+12 | AG Peg | 21 51 01.9747 | +12 37 32.137 | 9.0 | HIP 107848; slow nova |
| 2328+48 | Z And | 23 33 39.9565 | +48 49 06.001 | 10.8 | HIP 116287; class prototype; radio jet |
| 2136+02 | NSV 25735 | 21 41 44.8 | +02 43 54.4 | 10.5 | A&AS 146, 407 = StHa 190; few data points |
| 2328–15 | R Aqr | 23 43 49.4416 | –15 17 03.917 | 9.1 | HIP 117054; known jet source; Mira red giant |

Table 1. Symbiotic stars currently in the AAVSO International Database, continued.

*The AAVSO light curve for each object in this table may be found on the AAVSO light-curve-generator website: http://www.aavso.org/data/lcg/. Objects and V magnitudes taken from Belczyński et al. (2000) who state: "As most (if not all) symbiotics are variable, these values are arbitrary (usually the average of published measurements) just to give the general level of an object's brightness." For background on individual objects, see Belczyński et al. (2000) or Kenyon (1986).*

*Coordinates (except those from A&AS 146, 407) were assembled and verified by R. Webbink. They are drawn (in priority order) from* The Hipparcos *(HIP) and* Tycho *(TYC) catalogs (Perryman* et al. *1997),* U.S. Naval Observatory CCD Astrographic Catalog *(UCAC1) (Zacharias* et al. *2000),* Carlsberg Meridian Catalog *(CMC) (Copenhagen U. Obs.* et al. *1999), astrometry by Henden and Munari (2000, 2001), the* U.S. Naval Observatory A2.0 Catalog *(USNO-A2.0) (Monet,* et al. *1998),* Guide Star Catalog *1.1 (GSC) (Space Tel. Sci. Inst. 1992), and from R. Webbink's measurements in the GSC frame on the* Guide Star Survey *using the STScI GASP software (Space Tel. Sci. Inst. 2003).*

Table 2. X-Ray bright symbiotics soon to be added to the AAVSO International Database.

| Name | R. A. (2000)<br>h  m  s | Dec. (2000)<br>°  '  " | V | Coordinates Source / Comments |
|---|---|---|---|---|
| SMC 3 | 00 48 20.186 | −73 31 51.59 | 15.5 | GSS<br>= [HFP2000] 512 |
| LN 358 | 00 59 12.2548 | −75 05 17.686 | 15.2 | UCAC1 00750059<br>= SMC LN 358 |
| LMC S63 | 05 48 43.4158 | −67 36 10.264 | 15.2 | UCAC1 02294530<br>= AL 427 = V4435 LMC |
| Draco C-1 | 17 19 57.661 | +57 50 05.74 | 17 | A&AS 143, 343<br>= [ALS82] C1 = V225 Dra |
| Hen 3-1591 | 18 07 32.030 | −25 53 43.69 | 12.5 | GSC 6846-00296<br>= NSV 10219 |

*V magnitudes taken from Belczýnski et al. (2000) as described in Talbe 1 notes. Coordinates were assembled and verified by R. Webbink. Sources of coordinates are as described in Table 1 notes.*